\documentclass[a4paper,USenglish,cleveref, autoref, thm-restate]{lipics-v2021}

\pdfoutput=1 
\hideLIPIcs  

\usepackage{hyperref}

\usepackage{amsmath}
\usepackage{upgreek}
\usepackage{xcolor}
\usepackage{colortbl}
\usepackage[capitalise]{cleveref}

\usepackage{pbox}
\usepackage{makecell}

\usepackage{tikz-cd}
\usepackage{tabularx}

\definecolor{keywordcolor}{rgb}{0.7, 0.1, 0.1}   
\definecolor{tacticcolor}{rgb}{0.0, 0.1, 0.3}    
\definecolor{commentcolor}{rgb}{0.4, 0.4, 0.4}   
\definecolor{stringcolor}{rgb}{0.5, 0.3, 0.2}    
\definecolor{symbolcolor}{rgb}{0.1, 0.2, 0.7}    
\definecolor{sortcolor}{rgb}{0.1, 0.5, 0.1}      
\definecolor{attributecolor}{rgb}{0.7, 0.1, 0.1} 
\definecolor{errorcolor}{rgb}{1, 0, 0}           

\usepackage{listings}
\usepackage{flushend}
\usepackage{xspace}
\usepackage{underscore}
\usepackage{microtype}
\usepackage{datetime}
\usepackage{MnSymbol} 
\usepackage{ragged2e}

\newcommand{\ZZ}{\mathbb{Z}}

\newcommand{\RR}{\mathbb{R}}
\newcommand{\CC}{\mathbb{C}}

\newcommand{\mathlib}{\textsf{mathlib}\xspace}

\renewcommand{\phi}{\varphi}

\newcommand{\lean}[1]{\lstinline[language=lean]{#1}}

\usepackage{scalefnt}

\title{Formalized~functional~analysis~with~semilinear~maps}

\author{Fr\'ed\'eric Dupuis}
{Département d'informatique et de recherche opérationnelle, Université de Montréal, Canada \and \url{http://www.iro.umontreal.ca/~dupuisf/}}
{dupuisf@iro.umontreal.ca}
{https://orcid.org/0000-0002-5586-0505}
{} 

\author{Robert Y. Lewis}
{Department of Computer Science, Brown University, USA \and \url{https://robertylewis.com}}
{robert_lewis@brown.edu}
{https://orcid.org/0000-0002-5266-1121}
{}

\author{Heather Macbeth}
{Department of Mathematics, Fordham University, USA \and \url{https://faculty.fordham.edu/hmacbeth1/}}
{hmacbeth1@fordham.edu}
{https://orcid.org/0000-0002-0290-4172}
{} 

\authorrunning{F. Dupuis, R.\,Y. Lewis, and H.\,R. Macbeth} 

\Copyright{Fr\'ed\'eric Dupuis and Robert Y. Lewis and Heather Macbeth} 



\acknowledgements{We thank 
Johan Commelin for many conversations about isocrystals, and 
Johannes H\"olzl for comments on work in Isabelle.}

\nolinenumbers 

\EventEditors{John Q. Open and Joan R. Access}
\EventNoEds{2}
\EventLongTitle{42nd Conference on Very Important Topics (CVIT 2016)}
\EventShortTitle{CVIT 2016}
\EventAcronym{CVIT}
\EventYear{2016}
\EventDate{December 24--27, 2016}
\EventLocation{Little Whinging, United Kingdom}
\EventLogo{}
\SeriesVolume{42}
\ArticleNo{23}

\begin{document}

\maketitle

\begin{abstract}
Semilinear maps are a generalization of linear maps between vector spaces 
where we allow the scalar action to be twisted by a ring homomorphism
such as complex conjugation. 
In particular, this generalization unifies the concepts of linear and conjugate-linear maps.
We implement this generalization in Lean's \mathlib library,
along with a number of important results in functional analysis 
which  previously were impossible to formalize properly.
Specifically, we prove the Fr\'echet--Riesz representation theorem
and the spectral theorem for compact self-adjoint operators
generically over real and complex Hilbert spaces. 
We also show that semilinear maps have applications beyond functional analysis 
by formalizing the one-dimensional case of a theorem of Dieudonn\'e and Manin 
that classifies the isocrystals over an algebraically closed field
with positive characteristic.
\end{abstract}

\section{Introduction}
\label{sec:intro}

Proof assistant users have long recognized the value of abstraction. 
Working at high levels of generality
and specializing only when needed
can save significant effort in both the long and short term.
In program verification, this principle manifests 
in the use of stepwise refinement of programs 
from abstract specifications to executable code~\cite{Lamm19,Wirt71}.
Mathematical generalizations that are rarely used in informal presentations 
are much more common in formal libraries,
including the use of filters to generalize limits in topology and analysis~\cite{Holz13}
and uniform spaces as a generalization of metric spaces~\cite{Aff18,Bold15,BCM20}.

We propose another such mathematical generalization:
\emph{linear maps}, a fundamental concept in many fields of mathematics,
can be seen as a special case of \emph{semilinear maps}.
A linear algebra library built on top of this more general structure
can unify concepts that would otherwise be defined separately.
In particular, 
linear and \emph{conjugate-linear} (or \emph{antilinear}) maps are both examples of semilinear maps. 
By relating these, 
one can avoid a large amount of code duplication 
and state many theorems more naturally.
This generalization is rarely seen explicitly in informal mathematics.
Texts tend to focus on the linear case,
claiming results about the conjugate-linear or semilinear cases ``by analogy'' when needed. 

Motivated by the desire to formalize theorems from functional analysis 
at the proper level of abstraction,
we have implemented this generalization in \mathlib~\cite{mathlib20},
a library of formal mathematics in the Lean proof assistant~\cite{Mour15a}.
When we started this project, 
much of \mathlib was already built on top of standard linear maps.
With care and clever notation 
we were able to make the transition largely invisible.
With the generalization complete 
we were able to state and prove a number of theorems 
far more elegantly than could have been done before. 

Among the results unlocked by this refactor are 
the Fr\'echet--Riesz representation theorem,
which states that a Hilbert space is either isomorphic or conjugate-isomorphic to its dual space; 
the generic definition of the adjoint operator on an inner product space over $\RR$ or $\CC$;
and the spectral theorem for compact self-adjoint operators on Hilbert spaces,
which gives a canonical form for an important class of linear maps by reference to their eigenvectors.
This material in turn lays the groundwork for the formalization of vast areas of mathematics: 
complex Hilbert spaces are the bread and butter of quantum mechanics and are therefore
a prerequisite for quantum information theory and a large part of mathematical physics.

Furthermore, as evidence that semilinear maps are useful for more than unifying real and complex vector spaces,
we have also formalized the one-dimensional case of a theorem of Dieudonn\'e and Manin~\cite{Mani63}
that classifies the isocrystals over an algebraically closed field of characteristic $p>0$.
This is a foundational result in $p$-adic Hodge theory.

Related literature documents the struggles in other libraries 
to unify real and complex linear algebra. 
For instance, Aransay and Divas\'on~\cite{Aransay2017}, working in Isabelle, write:
\begin{quote}
  We miss \ldots the definition of a ``common place'' or generic structure 
  representing inner product spaces over real and complex numbers \ldots 
  that could permit a definition and formalisation of the Gram-Schmidt process 
  for both structures simultaneously.
\end{quote}
Their work introduces a ``local'' solution to the issue, 
but we argue that basing a library on semilinear maps is the ``global'' solution.
We discuss related work in more detail in Section~\ref{sec:conclusion}.

We estimate that over the course of this project we have added 13k lines of code to \mathlib, 
with 2k more lines waiting to be merged.
We provide links to our contributions, 
indicating where they can be found in the library,
on the project website.\footnote{
  \url{https://robertylewis.com/semilinear-paper}
}

\section{Mathematical preliminaries}

\label{sec:math-prelim}

\subsection{Semilinear maps}
Given modules $M_1$, $M_2$ over semirings $R_1$, $R_2$ and a ring homomorphism $\sigma:R_1\to R_2$, a \emph{$\sigma$-semilinear map} from $M_1$ to $M_2$ is a function $f:M_1\to M_2$ satisfying the two axioms
\begin{enumerate}
  \item for all $x, y\in M_1$, $f(x+y)=f(x)+f(y)$
  \item for all $x\in M_1$ and $c\in R_1$, $f(cx)=\sigma(c)f(x)$.
\end{enumerate}

Let us note the two canonical examples:
\begin{itemize}
\item For $R_1=R_2=R$ and $\sigma$ the identity ring homomorphism $\operatorname{id}_R:R\to R$, the second condition simplifies to $f(cx)=cf(x)$, and therefore an $\operatorname{id}_R$-semilinear map is precisely an $R$-linear map in the classic sense.

\item For $R_1=R_2=\mathbb{C}$ and $\sigma$ the complex-conjugation operation $\operatorname{conj}:\mathbb{C}\to\mathbb{C}$, the second condition simplifies to $f(cx)=\overline{c}f(x)$.  Therefore a $\operatorname{conj}$-semilinear map is a conjugate-linear map between complex vector spaces.
\end{itemize}

The theory of semilinear maps develops along the same lines as the theory of linear maps, with minimal adjustment.  The composition of a $\sigma$-semilinear map and a $\tau$-semilinear map, for $\sigma:R_1\to R_2$ and $\tau:R_2\to R_3$, is a $(\tau\circ\sigma)$-semilinear map.  (For example, the composition of two conjugate-linear maps is a linear map.)  If $\sigma$ is bijective, the inverse of a bijective $\sigma$-semilinear map is a $\sigma^{-1}$-semilinear map.

Theorems about special classes of linear maps also admit semilinear analogues.  Consider, for example, the theorem that a $\mathbb{K}$-linear map $f:E_1\to E_2$, for $\mathbb{K}$ a normed field and $E_1$, $E_2$ normed spaces over $\mathbb{K}$, is continuous if and only if it is \emph{bounded} ($\lVert f(x)\rVert\le M\lVert x\rVert$ for some fixed $M$, for all $x$).  This theorem generalizes to $\sigma$-semilinear maps, for $\sigma:\mathbb{K}_1\to \mathbb{K}_2$, if the ring homomorphism $\sigma$ is an isometry.

\subsection{Conjugate-linear maps in functional analysis}

An \emph{inner product space} is a vector space $E$ over a scalar field $\mathbb{K}\in\{\mathbb{R}, \mathbb{C}\}$ equipped with an \emph{inner product} $\langle\cdot, \cdot\rangle$, namely a $\mathbb{K}$-valued function of two arguments which is conjugate-linear in the first argument and linear in the second argument and which has symmetry and positivity properties:
\begin{enumerate}
  \item for all $u, v, w\in E$, $\langle u + v, w\rangle = \langle u, w\rangle + \langle  v, w\rangle$ and $\langle w, u + v\rangle = \langle w, u\rangle + \langle w, v\rangle$;
  \item for all $c\in\mathbb{K}$ and $v,w\in E$, $\langle cv, w\rangle = \overline{c}\langle v, w\rangle$ and $\langle v, cw\rangle = c\langle v,w\rangle$;
  \item for all $v, w\in E$, $\langle v, w\rangle = \overline{\langle w, v\rangle}$;
  \item for all $v\in E$, the quantity $\langle v, v\rangle$ (which by (3) is real) is nonnegative, and strictly positive unless $v=0$.
\end{enumerate}
For the case of real scalars, $\mathbb{K}=\mathbb{R}$, we consider the conjugation operation as being the identity; this allows a development of the complex case to subsume the simpler real case.

An inner product space has an associated norm $\lVert v\rVert=\sqrt{\langle v, v\rangle}$ and hence a metric structure.  A \emph{Hilbert space} is an inner product space for which this metric is complete.  This condition is automatic in finite dimension.

The \emph{dual} of an inner product space $E$ is the $\mathbb{K}$-vector space of continuous linear maps $\varphi:E\to \mathbb{K}$. 
There is a natural conjugate-linear map from $E$ to its dual $E^*$: the vector $v\in E$ is mapped to the vector $\langle v, \cdot \rangle$ in $E^*$.  To see the conjugate-linearity of this map, observe that $\langle cv, \cdot\rangle = \overline{c}\langle v, \cdot \rangle$.  It is not difficult to see that, for an appropriate norm on $E^*$, this map is an isometry. A more subtle theorem, the {\bf Fr\'echet--Riesz representation theorem}, asserts that for a Hilbert space $E$ this conjugate-linear map is bijective.

Given Hilbert spaces $E$, $F$ over $\mathbb{K}$ and a continuous linear map $T:E\to F$, it can be proved that there is a unique continuous linear map $T^*:F\to E$, the \emph{adjoint} of $T$, such that for all $v\in E$ and $w\in F$, $\langle Tv, w\rangle = \langle v, T^*w\rangle$. It turns out that the operation of sending $T:E\to F$ to its adjoint $T^*:F\to E$ is a conjugate-linear map from $E \to F$ to $F\to E$. To see the conjugate-linearity in this case, observe that 
\[
\langle v, (cT)^*w\rangle =\langle (cT)v, w\rangle=\overline{c}\langle Tv, w\rangle = \overline{c}\langle v, T^*w\rangle = \langle v, (\overline{c}T^*)w\rangle.
\]
Like the conjugate-linear map appearing in the Fr\'echet--Riesz representation theorem, the adjoint map $T \mapsto T^*$ turns out to be bijective and (for an appropriate norm) isometric.

Several important classes of continuous linear maps are defined using the adjoint.  A continuous linear map $T:E\to E$ is \emph{self-adjoint}, if $T^*=T$, and it is \emph{normal}, if $T^*T=TT^*$.  Self-adjoint implies normal.  

The \emph{Hilbert sum} $\bigoplus_{i\in\iota} E_\iota$ of a  family of inner product spaces $(E_i)_{i\in\iota}$ is an inner product space   whose elements are choices $(v_i)_{i\in\iota}$ of an element from each $E_i$, such that the collection of chosen elements is square-summable in the sense that $\sum_{i\in\iota}\lVert v_i\rVert^2<\infty$.  Elements in the Hilbert sum $\bigoplus_{i\in\iota} E_i$ can be added and scalar-multiplied in the obvious way.  The inner product on the Hilbert sum is given by $\langle (v_i)_{i\in\iota}, (w_i)_{i\in\iota}\rangle = \sum_{i\in\iota}\langle v_i, w_i\rangle$.  It can be proved that if each $E_i$ is a Hilbert space (i.e., complete) then so is $\bigoplus_{i\in\iota} E_i$.  A linear map 
$T:\bigoplus_{i\in\iota} E_i \to \bigoplus_{i\in\iota} E_i$
is \emph{diagonal} if there exist scalars $(\mu_i)_{i\in\iota}$ such that for all $(v_i)_{i\in\iota}\in \bigoplus_{i\in\iota} E_i$, 
$T\left((v_i)_{i\in\iota}\right)=(\mu_i v_i)_{i\in\iota}$.

A linear map $T:E\to F$ between normed spaces is \emph{compact} if the image under $T$ of the unit ball in $E$ is precompact (that is, has compact closure) in $F$.  This condition implies the continuity of $T$ but is more restrictive.  The {\bf spectral theorem} states that a normal (over $\mathbb{C}$) or self-adjoint (over $\mathbb{R}$ or $\mathbb{C}$), compact linear map $T:E\to E$ is equivalent to a diagonal map, in the sense that there exists a bijective linear isometry $\Phi$ from $E$ to a Hilbert sum $\bigoplus_{i\in\iota} F_i$, such that the linear map 
$\Phi\circ T\circ \Phi^{-1}$ is diagonal. In fact, the $F_i$ may be chosen to be the eigenspaces of $T$, with the $\mu_i$ chosen to be the associated eigenvalues.

In finite dimension, every linear map is compact.  In this setting the spectral theorem reduces to the more elementary {\bf diagonalization theorem} for a normal endomorphism $T$ of a finite-dimensional inner product space $E$: there exists a bijective linear isometry $\Phi$ from $E$ to a finite direct sum of finite-dimensional inner product spaces $(F_i)_{i\in\iota}$, such that the linear map $\Phi\circ T\circ \Phi^{-1}$ is diagonal.

\subsection{Frobenius-semilinear maps}
\label{subsec:math-prelim-frobenius}

Given a commutative ring $R$ and a prime $p$, there is a classical construction~\cite{Haze09} of an associated commutative ring $\mathbb{W}(R)$, the ring of \emph{$p$-typical Witt vectors} of $R$.  The elements of this ring are sequences of elements of $R$, but the definitions of addition and multiplication are rather elaborate.  The motivating example is that for $R$ the finite field $\mathbb{Z}/p\mathbb{Z}$, the ring $\mathbb{W}(\mathbb{Z}/p\mathbb{Z})$ is the ring of $p$-adic integers.

The ring $\mathbb{W}(R)$ admits a canonical ring-endomorphism, the \emph{Frobenius endomorphism}.  Concretely, it operates by sending a sequence $(x_0, x_1, x_2, \ldots)$ to $(x_0^p, x_1^p, x_2^p,\ldots)$.  In the example of the $p$-adic integers  $\mathbb{W}(\mathbb{Z}/p\mathbb{Z})$, this endomorphism is the identity, so the construction becomes interesting only for more complicated rings $R$, such as field extensions of $\mathbb{Z}/p\mathbb{Z}$.

For $R$ an integral domain of characteristic $p$, the ring $\mathbb{W}(R)$ is also an integral domain, and therefore has a well-defined field of fractions.  In this case, the Frobenius endomorphism of  $\mathbb{W}(R)$ extends to an endomorphism of its field of fractions.  If moreover the ring $R$ is perfect, then the Frobenius endomorphism is an automorphism (that is, bijective), as is the induced automorphism of its field of fractions.

Let us fix an algebraically closed field $R$ of characteristic $p$ (which is necessarily a perfect integral domain), and denote by $K$ the field of fractions of $\mathbb{W}(R)$ and by $\varphi:K\to K$ the Frobenius automorphism of $K$.  There is a very well-developed theory of $\varphi$-semilinear maps between vector spaces over $K$.  Notably, an important {\bf theorem of Dieudonn\'e and Manin}~\cite{Mani63} provides an analogue of the spectral theorem.  For a finite-dimensional vector space $V$ over $K$, it classifies the \emph{isocrystals} (bijective $\varphi$-semilinear maps $f:V \to V$), by constructing for such an $f$ a decomposition of $V$ as a direct sum of vector spaces $V_i$ which are preserved by $f$ and on each of which the map $f$ has a certain canonical form.

\section{Lean preliminaries}
\label{sec:lean-prelim}

The \mathlib library builds its algebraic hierarchy using \emph{type classes}~\cite{mathlib20,Spit11}.
Baanen~\cite{Baanen2022} gives an in depth account of \mathlib's use of type classes,
which we summarize very briefly.

Each argument to a Lean declaration is declared as 
\emph{explicit} (\lstinline{()}), \emph{implicit} (\texttt{\{\}}), or \emph{instance-implicit} (\lstinline{[]}).
Explicit arguments must be provided when the declaration is applied;
implicit arguments are inferred by unification;
instance-implicit arguments are inferred by type class instance resolution.

The fundamental type class of \mathlib's linear algebra library is \lean{module}. 
\begin{lstlisting}
class module (R : Type u) (M : Type v) [semiring R] [add_comm_monoid M] 
  extends distrib_mul_action R M :=
(add_smul : ∀ (r s : R) (x : M), (r + s) • x = r • x + s • x)
(zero_smul : ∀ (x : M), (0 : R) • x = 0)
\end{lstlisting}
This type class says that the additive monoid \lean{M} has an \lean{R}-module structure:
it supports scalar multiplication by elements of the semiring \lean{R},
and this scalar multiplication behaves properly with respect to addition on \lean{M}.
When \lean{R} is a field instead of a semiring, 
an \lean{R}-module is in fact a vector space. 
Many definitions and theorems apply in the more general setting,
and when the vector space setting is needed,
the transition is invisible. 

A type class is a structure (i.e. a record type) 
that takes zero or more \emph{parameters} and has zero or more \emph{fields}.
In the above, the arguments \lean{R} and \lean{M} are parameters,
as are the arguments that \lean{R} is a semiring and \lean{M} is an additive commutative monoid. 
In order to elaborate the type \lean{module R M},
Lean's type class inference algorithm must be able to infer the latter arguments automatically. 
The fields of \lean{module} are \lean{add_smul} and \lean{zero_smul},
and a projection to \lean{distrib_mul_action R M}.
To construct a term of type \lean{module R M}, the user must provide these values;
given a term of type \lean{module R M}, the user can access these values. 
The \lean{extends} keyword can be read as ``inherits from.''
An assumption \lean{distrib_mul_action R M} is available while defining the fields \lean{add_smul} and \lean{zero_smul},
and indeed, the scalar action used in these fields is derived from this instance.

By default the parameters to a type class are \emph{input parameters}.
Lean will begin its instance search when all input parameters are known.
By denoting certain parameters as \emph{output parameters},
the user can instruct Lean to begin searching for instances of that class 
before those parameters are known;
they will be determined by the solution to the search.
Baanen~\cite[Section 5.1]{Baanen2022} describes output parameters in more detail.

Like \mathlib, we freely use classical logic and do not focus on defining things computably. 
Within code blocks in this paper, 
we omit the bodies of definitions and theorems 
when only the type is relevant,
omit some implicit arguments when the types are clear from context,
and occasionally rename declarations for the sake of presentation.

\section{Semilinear maps in Lean}
\label{sec:semilin-maps}

Section~\ref{sec:math-prelim} covered the mathematical motivation for semilinear maps. 
Here we focus on our implementation of this generalization in Lean. 
This work is done in the context of \mathlib~\cite{mathlib20},
a project with over 780k lines of code, 220 contributors, and countless users.
Given the difficulty and importance of maintaining such a large library~\cite{DEL20},
we were motivated to make this refactor with as little disruption as possible.

\subsection{Defining semilinear maps}
\label{subsec:semilin-def}

Before beginning our refactor to use semilinear maps,
\mathlib's linear algebra development was based on the more familiar concept of linear maps. 

\begin{lstlisting}
structure linear_map (R : Type u) (M₁ : Type v) (M₂ : Type w)
  [semiring R] [add_comm_monoid M₁] [add_comm_monoid M₂] [module R M₁]
  [module R M₂] extends add_hom M₁ M₂, mul_action_hom R M₁ M₂
\end{lstlisting}

Given two \lean{R}-modules \lean{M₁} and \lean{M₂},
a linear map is an additive homomorphism \lean{M₁ → M₂} 
that respects the multiplicative action of \lean{R}.
A \lean{mul_action_hom} is a homomorphism 
between types acted on by the same type of scalars~\cite{Wieser2021}.

For readers unfamiliar with Lean syntax,
it may be clarifying to see what information goes in to defining such a linear map.
Despite the intimidating syntax,
the input information is exactly as expected:
if you have types \lean{R}, \lean{M₁}, and \lean{M₂} with the appropriate operations and structure,
you can construct a linear map by providing
a function \lean{M₁ → M₂} and 
proofs that this function factors through addition and scalar multiplication.

\begin{lstlisting}
example (R : Type u) (M₁ : Type v) (M₂ : Type w)
  [semiring R] [add_comm_monoid M₁] [add_comm_monoid M₂] [module R M₁]
  [module R M₂] : linear_map R M₁ M₂ := 
{ to_fun := _, -- M₁ → M₂
  map_add' := _, -- ∀ (x y : M₁), to_fun (x + y) = to_fun x + to_fun y
  map_smul' := _ } -- ∀ (m : R) (x : M₁), to_fun (m • x) = m • to_fun x
\end{lstlisting}

As noted in Section~\ref{sec:math-prelim},
the domain and codomain of a linear map are modules over the same semiring. 
The same is true in the definition of linear equivalences:

\begin{lstlisting}
structure linear_equiv (R : Type u) (M₁ : Type v) (M₂ : Type w)
  [semiring R] [add_comm_monoid M₁] [add_comm_monoid M₂] [module R M₁]
  [module R M₂] extends linear_map R M₁ M₂, add_equiv M₁ M₂
\end{lstlisting}

The type signature of a semilinear map\footnote{
  In our \mathlib contribution 
  we did not rename the type \lean{linear_map} to \lean{semilinear_map}.
  This simplified the refactor and makes the definition easier to find for beginners. 
  For the sake of clarity in this paper, 
  we refer to the generalized type by the more accurate name.
} is more complicated, 
involving two scalar semirings and a ring homomorphism between them.
It no longer makes sense to extend \lean{mul_action_hom},
since the multiplicative actions are over different scalar types, 
so we instead add the field \lean{map_smul} directly.
The arguments \lean{R} and \lean{S} can be inferred from \lstinline{σ}
and are thus marked as implicit.
The type \lean{R →+* S} is the type of ring homomorphisms from \lean{R} to \lean{S}.

\begin{lstlisting}
structure semilinear_map {R : Type*} {S : Type*} [semiring R] [semiring S] 
  (σ : R →+* S) (M₁ : Type*) (M₂ : Type*)
  [add_comm_monoid M₁] [add_comm_monoid M₂] [module R M₁] [module S M₂] 
  extends add_hom M₁ M₂ :=
(map_smul' : ∀ (r : R) (x : M₁), to_fun (r • x) = (σ r) • to_fun x)
\end{lstlisting}

While the type signature has grown more complicated,
the constructor for a semilinear map is quite similar 
to that of a linear map:

\begin{lstlisting}
example {R : Type*} {S : Type*} [semiring R] [semiring S] 
  (σ : R →+* S) (M₁ : Type*) (M₂ : Type*)
  [add_comm_monoid M₁] [add_comm_monoid M₂] [module R M₁] [module S M₂] : 
  semilinear_map σ M₁ M₂ :=
{ to_fun := _, -- M₁ → M₂
  map_add' := _, -- ∀ (x y : M₁), to_fun (x + y) = to_fun x + to_fun y
  map_smul' := _ } -- ∀ (r : R) (x : M₁), to_fun (r • x) = σ r • to_fun x
\end{lstlisting}

The generalization to semilinear equivalences is similar, but more involved in order to gracefully handle inversion of such maps. The additional parameter \lstinline{σ'} and the \lean{ring_hom_inv_pair} type class are explained in Section~\ref{subsec:composition}.
\begin{lstlisting}
structure semilinear_equiv {R : Type*} {S : Type*} [semiring R] [semiring S] 
  (σ : R →+* S) {σ' : S →+* R} [ring_hom_inv_pair σ σ'] 
  [ring_hom_inv_pair σ' σ] (M₁ : Type*) (M₂ : Type*)
  [add_comm_monoid M₁] [add_comm_monoid M₂] [module R M₁] [module S M₂]
  extends linear_map σ M₁ M₂, add_equiv M₁ M₂
\end{lstlisting}

\subsection{Notation for semilinear maps}
\label{subsec:notation}

One can see from these definitions that 
semilinear maps are not a drop-in replacement for linear maps. 
The type signature is different, even when looking only at explicit arguments. 
To convert an \lean{R}-linear map to a semilinear map, 
one must know to invoke \lean{ring_hom.id R},
the identity ring homomorphism on \lean{R}.

Given how frequently linear maps appear in \mathlib, 
this refactor threatened to be painful. 
Our job was made immensely easier by the use of notation. 
Before our refactor 
\mathlib used the notation \lean{M₁ →ₗ[R] M₂} to stand for for \lean{linear_map R M₁ M₂}.
By redefining this notation to stand for \lean{semilinear_map (ring_hom.id R) M₁ M₂}
we were largely able to avoid breaking definitions and proofs throughout the library. 
The same approach, with notation \lean{M₁ ≃ₗ[R] M₂},
worked to generalize linear equivalences to semilinear equivalences.
We introduced similar notation \lean{M₁ →ₛₗ[σ] M₂} to stand for \lean{semilinear_map σ M₁ M₂},
and \lean{M₁ →ₗ⋆[R] M₂} to stand for 
a semilinear map with respect to a fixed involution such as complex conjugation.

The composition of linear maps 
proved to be a complication.
As we note in Section~\ref{subsec:composition},
an additional type class must be inferred 
to justify that two semilinear maps can be composed.
This inference was fragile in the presence of other features, like implicit coercions, that complicate elaboration.
We introduced notation \lstinline{∘ₗ} for the composition of linear maps,
using \lean{ring_hom.id} to justify the composition, 
and manually inserted this notation where needed.

For our new definition to be useful,
theorems stated for linear maps \lean{M₁ →ₗ[R] M₂} needed to be upgraded 
to theorems about semilinear maps \lean{M₁ →ₛₗ[σ] M₂} when possible. 
Doing so is mostly mechanical
and our use of notation let us approach this without hurry.
Because theorems generalized to semilinear maps still apply directly to the linear case
we were able to do this generalization incrementally from the bottom up. In particular,
several more specialized classes of linear maps and equivalences are also present in \mathlib
(Figure~\ref{figure:map-table}).
Our bottom-up approach allowed us to 
generalize these one at a time.

\begin{figure}
  {\RaggedRight
  \begin{tabularx}{\textwidth}{p{1.8cm}|l|l|l|X}
    & Linear & Conjugate-linear & Semilinear  & Meaning \\ \hline
    Map & \lean{M₁ →ₗ[R] M₂} & \lean{M₁ →ₗ⋆[R] M₂} & \lean{M₁ →ₛₗ[σ] M₂} & Between modules; factors over addition and scalar multiplication \\ \hline 
    Continuous map & \lean{M₁ →L[R] M₂} & \lean{M₁ →L⋆[R] M₂} & \lean{M₁ →SL[σ] M₂} & Between topological modules; a continuous map \\ \hline
    Equivalence & \lean{M₁ ≃ₗ[R] M₂} & \lean{M₁ ≃ₗ⋆[R] M₂} & \lean{M₁ ≃ₛₗ[σ] M₂} & An invertible map \\ \hline 
    Isometry & \lean{M₁ ≃ₗᵢ[R] M₂} & \lean{M₁ ≃ₗᵢ⋆[R] M₂} & \lean{M₁ ≃ₛₗᵢ[σ] M₂} & Between normed modules; a norm-preserving equivalence 
  \end{tabularx}}

  \caption{Notation for various classes of (semi)linear operators that appear in this paper}
  \label{figure:map-table}
\end{figure}

\subsection{Composition of semilinear maps}
\label{subsec:composition}

Composition of maps is complicated by this generalization.
The composition of two linear maps is straightforward:
it is easy to check that the composition of the underlying functions
preserves addition and scalar multiplication. 
With semilinear maps one must also compose the homomorphisms between scalar rings.
Given \lean{f : M₁ →ₛₗ[σ₁₂] M₂} and \lean{g : M₂ →ₛₗ[σ₂₃] M₃}, 
we would naturally end up with \lean{g.comp f : M₁ →ₛₗ[σ₂₃.comp σ₁₂] M₃}.

This ends up being awkward to handle in many common situations. 
Suppose we wish to state that 
\lean{f : M₁ →ₛₗ[σ₁₂] M₂} and \lean{g : M₂ →ₛₗ[σ₂₁] M₁} are inverse maps:
\lean{f.comp g = (id : M₁ →ₗ[R] M₁)}.
This statement is not type-correct, 
since the ring homomorphism on the left is \lstinline{σ₁₂.comp σ₂₁} 
and the one on the right is the identity. 
Such an issue appears in practice, for example, 
when defining the adjoint as a conjugate-linear map (Section~\ref{sec:adjoints}).

To solve this issue, 
we introduce a type class \lean{ring_hom_comp_triple}
that states that two ring homomorphisms compose to a third. 
\begin{lstlisting}
class ring_hom_comp_triple [semiring R₁] [semiring R₂] [semiring R₃]
  (σ₁₂ : R₁ →+* R₂) (σ₂₃ : R₂ →+* R₃) (σ₁₃ : out_param (R₁ →+* R₃)) : Prop :=
(comp_eq : σ₂₃.comp σ₁₂ = σ₁₃)
\end{lstlisting}

We register a number of global instances of this class.  We then use the \lean{ring_hom_comp_} \lean{triple} type class in the definition of composition.
\begin{lstlisting}
def semilinear_map.comp {R₁ R₂ R₃ : Type*} {M₁ M₂ M₃ : Type*}
  [semiring R₁] [semiring R₂] [semiring R₃] 
  [add_comm_monoid M₁] [add_comm_monoid M₂] [add_comm_monoid M₃] 
  {mod_M₁ : module R₁ M₁} {mod_M₂ : module R₂ M₂} {mod_M₃ : module R₃ M₃} 
  {σ₁₂ : R₁ →+* R₂} {σ₂₃ : R₂ →+* R₃} {σ₁₃ : R₁ →+* R₃} 
  [ring_hom_comp_triple σ₁₂ σ₂₃ σ₁₃] 
  (g : M₂ →ₛₗ[σ₂₃] M₃) (f : M₁ →ₛₗ[σ₁₂] M₂) : 
  (M₁ →ₛₗ[σ₁₃] M₃)
\end{lstlisting}

While this may appear to be a rather verbose type signature 
for the composition of maps, 
it allows us to avoid the above problem without introducing further complications. 
In common situations, 
the appropriate global instances generate the necessary \lean{ring_hom_comp_triple} argument
without input from the user. For example, the following global instance allows for the composition of two (genuine) linear maps, or more generally for the composition of a semilinear map with a linear map.
\begin{lstlisting}
instance [semiring R₁] [semiring R₂] {σ₁₂ : R₁ →+* R₂} : 
  ring_hom_comp_triple (ring_hom.id R₁) σ₁₂ σ₁₂ 
\end{lstlisting}
Another instance helps in the setting of conjugate-linear maps.\footnote{In fact, we do not
state this instance explicitly; it is derived by type class inference from the \lean{ring_hom_inv_pair} instance for \lean{conj} (see below) and yet another global instance generating a
\lean{ring_hom_comp_triple} with the identity from a \lean{ring_hom_inv_pair}.}
\begin{lstlisting}
instance [comm_semiring R] [star_ring R] :
  ring_hom_comp_triple (conj R) (conj R) (ring_hom.id R)
\end{lstlisting}
We expand on the types here in Section~\ref{subsec:r-or-c};
in concrete terms, this instance says that 
the conjugation operation on a type supporting conjugation is an involution.
This allows us to compose two conjugate-linear maps to obtain, definitionally, a linear map.
The intention is that users should never work directly with a composition 
\lean{g.comp f : M₁ →ₛₗ[σ₂₃.comp σ₁₂] M₃},
but instead with \lean{g.comp f : M₁ →ₛₗ[σ₁₃] M₃} 
for some \lstinline{σ₁₃} satisfying \lean{ring_hom_comp_triple σ₁₂ σ₂₃ σ₁₃},
which is strictly more general.

Similar issues appear with semilinear equivalences,
specifically when defining the symmetric equivalence:
if \lean{e : E ≃ₛₗ[σ] F}, 
the ``natural'' definition of the symmetric equivalence
would give \lean{e.symm : F ≃ₛₗ[σ.symm] E}.
Some ring homomorphisms, notably conjugation on $\CC$, 
have the property that \lstinline{σ.symm = σ}. 
But these equalities are rarely definitional 
and spurious \lean{symm}s can block type checking. 
Introducing a new type class \lean{ring_hom_inv_pair} that states that two ring homomorphisms are inverses of each other, analogous to the type class \lean{ring_hom_comp_triple} described above,
again solves this issue.
\begin{lstlisting}
class ring_hom_inv_pair [semiring R₁] [semiring R₂] (σ : R₁ →+* R₂)
  (τ : out_param (R₂ →+* R₁)) : Prop :=
(comp_eq : τ.comp σ = ring_hom.id R₁)
(comp_eq₂ : σ.comp τ = ring_hom.id R₂)
\end{lstlisting}
Now, with a suitable instance stating that the conjugation operation on a type supporting conjugation is its own inverse, we can work with a conjugate-linear equivalence \lean{e : E ≃ₗ⋆[R] F}, i.e. \lean{e : E ≃ₛₗ[conj] F}, over scalars of that type,
and have that its inverse \lean{e.symm} be genuinely of type \lstinline{F ≃ₗ⋆[R] E}.
\begin{lstlisting}
instance [comm_semiring R] [star_ring R] : ring_hom_inv_pair (conj R) (conj R) 
\end{lstlisting}    

\section{Fréchet--Riesz representation theorem}
\label{sec:frechet} 

In the following three sections we describe results 
that we were able to formalize at the proper level of generality
thanks to our refactor.
By the ``proper level'' of generality,
we mean that our results hold generically over the real and complex numbers 
without case splits.

\subsection{The \texttt{is\_R\_or\_C} type class}
\label{subsec:r-or-c}

Many results in functional analysis, including those presented here,
hold for a field $\mathbb{K} \in \{ \RR, \CC \}$. 
Such results are usually presented in the literature 
by giving proofs for the complex case, 
with the real case following in the obvious way: 
replace complex conjugation by the identity, $i$ by zero, and so on. 

Before beginning our refactor,
we introduced a type class \lean{is_R_or_C} to \mathlib 
used to formalize this kind of result.
A type that instantiates \lean{is_R_or_C}
is a complete nondiscrete field with (real) norm containing an element $i$ and functions 
\lean{conj}, \lean{re} and \lean{im} 
that satisfy a number of ad-hoc axioms 
chosen to mimic the behavior of a field that is either $\mathbb{R}$ or $\mathbb{C}$.
The \lean{conj} operator is an involutive ring homomorphism,
enabling the notation discussed in Section~\ref{subsec:composition}.
Two global instances stating \lean{is_R_or_C ℝ} and \lean{is_R_or_C ℂ}
allow theorems over the generic type class 
to be specialized immediately to either concrete type.
The conjugation operator \lean{conj}
is definitionally equal to the identity function in the real case 
and the complex conjugate in the complex case. 
We note an experiment with a similar type class in Isabelle~\cite{Aransay2017}.

Hilbert spaces in \mathlib are defined over \lean{is_R_or_C} fields. 
Given two Hilbert spaces \lean{E} and \lean{F} over a field \lean{K}, 
conjugate-linear maps \lean{E ≃ₗ⋆[K] F} are precisely maps which are semilinear with respect to \lean{conj},
and thus in the real case are linear maps by definition.
Within \mathlib, this type class has already been used
extensively beyond the results mentioned in this paper, notably by Sébastien Gouëzel for stating in correct generality the 
Hahn--Banach theorem,
the smooth case of the inverse function theorem, 
and more.

\subsection{Fréchet--Riesz representation theorem}
\label{subsec:frechet}
Our first application of semilinear maps is in proving the Fréchet--Riesz representation theorem.
While the real case has been formalized in Coq~\cite{Boldo17} and Mizar~\cite{Narita2015},
and the complex case in Isabelle~\cite{Caballero2021},
we are not aware of a development that unifies the two.\footnote{
This theorem should not be confused with the Riesz--Markov--Kakutani representation theorem,
which has been formalized in Mizar, PVS (unfortunately referred to as the ``Riesz representation theorem''), 
and possibly other proof assistants.
}

Given a Hilbert space $E$, 
its \emph{dual space} $E^*$ consists of 
the set of continuous linear functionals on $E$ 
(i.e. $E^* = \{ f : E \rightarrow \mathbb{K} \mid f \text{ is linear and continuous} \}$). 
The dual space certainly includes elements of the form $f_v$ 
that map $w \in E$ to $\langle v, w \rangle$, 
and the Fréchet--Riesz representation theorem states that 
all elements of the dual space are of this form. 
That is, 
there exists an (in fact, isometric) equivalence between $E$ and $E^*$
that maps $v$ to to $f_v$.

The difficulty in formalizing this is that 
while this equivalence is linear in the real case, 
in the complex case, it is \emph{conjugate}-linear. 
The challenge is to construct this object in such a way that 
(1) there is a common definition for both the real and complex case, 
and (2) the added complication of conjugate-linearity is completely transparent in the real case. 
Before our refactor \mathlib simply had two separate constructions.
We are able to replace those two constructions with the following, 
which satisfies both requirements stated above:
\begin{lstlisting}
def to_dual [is_R_or_C 𝕜] [inner_product_space 𝕜 E] [complete_space E] : 
  E ≃ₗᵢ⋆[𝕜] normed_space.dual 𝕜 E
\end{lstlisting}
\begin{lstlisting}
lemma to_dual_apply [is_R_or_C 𝕜] [inner_product_space 𝕜 E] 
  [complete_space E] {x y : E} : to_dual 𝕜 E x y = ⟪x, y⟫
\end{lstlisting}
  
Read aloud this definition says that 
``a real or complex Hilbert space $E$ is isometrically conjugate-isomorphic to its dual space.''
But when specialized to the real case, 
the statement is definitionally equal to 
``$E$ is isometrically isomorphic to its dual space.''

Our proof of this theorem does not differ from the real version of the proof 
in \mathlib prior to our refactor. 
In fact, the patch unifying the real and complex versions\footnote{
  \url{https://github.com/leanprover-community/mathlib/pull/9924}
}
added only 45 lines of code and removed 79; 
the only change beyond rearranging and documentation 
was to generalize the statement of the theorem.
The Lean implementation of the orthogonal projection on real inner product spaces,
a tool used in the proof, 
had been written by Zhouhang Zhou as a port of work in Coq by Boldo et al.~\cite{Boldo17}. 

\section{Adjoints of operators on Hilbert spaces}
\label{sec:adjoints}

Given a continuous linear map $A$ between two Hilbert spaces $E$ and $F$, the adjoint of $A$ is the unique continuous linear map $A^* : F \rightarrow E$ such that for all $x \in E$ and $y \in F$, $\langle y, A x \rangle_F = \langle A^* y, x \rangle_E$.
The adjoint satisfies a number of properties: it is involutive (i.e.~$(A^*)^* = A$), it is an isometry, and, most importantly for our purposes here, it is conjugate-linear. Hence, it was natural to bundle it in \mathlib{} as a conjugate-linear isometric equivalence as follows:
\begin{lstlisting}
def continuous_linear_map.adjoint [is_R_or_C 𝕜] [inner_product_space 𝕜 E]
  [inner_product_space 𝕜 F] [complete_space E] [complete_space F] : 
  (E →L[𝕜] F) ≃ₗᵢ⋆[𝕜] (F →L[𝕜] E)
\end{lstlisting}
\begin{lstlisting}
lemma continuous_linear_map.adjoint_inner_left [is_R_or_C 𝕜]
  [inner_product_space 𝕜 E] [inner_product_space 𝕜 F] [complete_space E]
  [complete_space F] (A : E →L[𝕜] F) (x : E) (y : F) :
  ⟪continuous_linear_map.adjoint A y, x⟫ = ⟪y, A x⟫
\end{lstlisting}

This definition fully exploits the algebraic formalism built for semilinear maps, including the composition mechanism of Section~\ref{subsec:composition}.  For example, the statement that the composition of the adjoint operation with itself is equal to the identity map from \lean{E →L[𝕜] F} to itself (a ``true'' \lstinline{𝕜}-linear map) would not typecheck without the \lean{ring_hom_comp_triple} mechanism.

In finite dimension, every linear map is a continuous linear map, so the adjoint construction actually applies
to every linear map.  We provide this construction as \lean{linear_map.adjoint}
for the benefit of future users interested only in the finite-dimensional setting.

An operator $T$ on a Hilbert space is said to be \emph{self-adjoint} if $T=T^*$ and \emph{normal} if $T^*T=TT^*$.  We allow these definitions to apply both to the finite-dimensional setting with \lean{linear_map.adjoint} and to the general setting with \lean{continuous_linear_map.adjoint} by in fact writing these definitions in the more general context of a \lean{star_ring}, a ring equipped with a fixed involutive ring homomorphism.

\begin{lstlisting}
def self_adjoint [ring R] [star_ring R] : add_subgroup R :=
{ carrier := {x | star x = x},   ... }
\end{lstlisting}
\begin{lstlisting}
def is_star_normal [ring R] [star_ring R] (x : R) :=
star x * x = x * star x
\end{lstlisting}

When \lean{R} is the ring \lean{E →ₗ[𝕜] E} of linear endomorphisms of a finite-dimensional inner product space \lean{E} (with ring operation composition), the involution \lean{star} is \lean{linear_map.adjoint}.
When \lean{R} is the ring \lean{E →L[𝕜] E} of continuous linear endomorphisms of a Hilbert space \lean{E}, the involution \lean{star} is \lean{continuous_linear_map.adjoint}.

\section{Versions of the spectral theorem}
\label{sec:spectral}

\subsection{The \texorpdfstring{$\ell^2$}{ell-two} construction}\label{subsec:hilbert-sum}

The spectral theorem, in finite dimension also known as the diagonalization theorem, expresses an operator on a Hilbert space in the canonical form of a ``diagonal'' operator.  To describe this canonical form, one needs some version of the \emph{Hilbert sum} or \emph{$\ell^2$-space} constructions.
Before we started, \mathlib already had a finitary version of this construction, namely the following construction for an inner product space structure on the product of finitely many inner product spaces.

\begin{lstlisting}
def pi_Lp {ι : Type u} (p : ℝ) (G : ι → Type v) : Type (max u v) :=
Π (i : ι), G i  
\end{lstlisting}
\begin{lstlisting}
instance pi_Lp.inner_product_space {𝕜 : Type w} [is_R_or_C 𝕜]
  {ι : Type u} [fintype ι] (G : ι → Type v)
  [Π (i : ι), inner_product_space 𝕜 (G i)] :
  inner_product_space 𝕜 (pi_Lp 2 G)
\end{lstlisting}  

We require the general version of this construction, with a possibly-infinite index set $\iota$.  We first define a predicate \lean{mem_ℓp f p} on dependent functions in \lstinline{Π (i : ι), G i} which, for $p=2$,
amounts to the norm-squared of the function being a convergent sum.
The associated subset of \lstinline{Π (i : ι), G i} is named \lean{lp G p}, proved to be an additive subgroup, and for $p=2$ equipped with an inner product space structure.
This inner product space is called the Hilbert sum of the family \lean{G}.

\begin{lstlisting}
def lp {ι : Type u} (G : ι → Type v) [Π (i : ι), normed_group (G i)]
  (p : ℝ≥0∞) : add_subgroup (Π (i : ι), G i) :=
{ carrier := {f | mem_ℓp f p},   ... }
\end{lstlisting}

\begin{lstlisting}
instance lp.inner_product_space {ι : Type u} {𝕜 : Type w} [is_R_or_C 𝕜]
  {G : ι → Type v} [Π (i : ι), inner_product_space 𝕜 (G i)] :
  inner_product_space 𝕜 (lp G 2)  
\end{lstlisting}

This is a reasonably labor-intensive construction (some 500 lines of code), 
the difficulties being a series of small analytic arguments 
about the convergence of the sums involved. 
It is closely analogous to R\'emy Degenne's \mathlib
construction of the inner product space structure on $L^2(X, G)$, 
with related work in Isabelle~\cite{Lp-AFP}.  
However, neither construction is a strict generalization of the other: 
the $L^2$ construction allows for integrals with respect to an arbitrary measure rather than just sums, 
whereas the $\ell^2$ construction applies to dependent functions of type \lstinline{Π (i : ι), G i} 
in which the ``codomain'' varies depending on the argument. 
We in fact need this dependent property for the spectral theorem.

A further analytic argument establishes the completeness of $\ell^p$.  The key step here is an argument that a pointwise limit of a uniformly-bounded sequence of elements of $\ell^p$ is itself in $\ell^p$. is  A Hilbert space is by definition a complete inner product space and therefore this establishes that the Hilbert sum \lean{lp G 2} is a Hilbert space.
\begin{lstlisting}
instance lp.complete_space {ι : Type u} {G : ι → Type v}
  [Π (i : ι), normed_group (G i)] [∀ (i : ι), complete_space (G i)] 
  {p : ℝ≥0∞} [fact (1 ≤ p)] : complete_space (lp G p)
\end{lstlisting}

Finally, given a Hilbert space $E$ of interest, an important argument establishes a mechanism for ``collating'' a family of isometries from the summands \lean{G i} into $E$ to an isometric isomorphism from \lean{lp G 2} into $E$. It is sufficient (and necessary) that the images of the family of isometries form a mutually-orthogonal family of subspaces of $E$, and that their joint span be dense in $E$.
\begin{lstlisting}
def orthogonal_family.linear_isometry_equiv [is_R_or_C 𝕜]
  [inner_product_space 𝕜 E] [complete_space E]
  [Π (i : ι), inner_product_space 𝕜 (G i)] {V : Π (i : ι), G i →ₗᵢ[𝕜] E}
  (hV : orthogonal_family 𝕜 V) [∀ (i : ι), complete_space (G i)]
  (hV' : (⨆ (i : ι), (V i).to_linear_map.range).topological_closure = ⊤) :
  E ≃ₗᵢ[𝕜] (lp G 2)
\end{lstlisting}  

We also provide the finitary, i.e. \lean{pi_Lp}, version of this construction. 
\begin{lstlisting}
def isometry_L2_of_orthogonal_family
  [is_R_or_C 𝕜] [inner_product_space 𝕜 E] [fintype ι] [decidable_eq ι]
  {V : ι → submodule 𝕜 E} (hV : direct_sum.submodule_is_internal V)
  (hV' : orthogonal_family 𝕜 (λ (i : ι), (V i).subtypeₗᵢ)) :
  E ≃ₗᵢ[𝕜] pi_Lp 2 (λ (i : ι), V i)
\end{lstlisting}  

\subsection{Common outline of the spectral theorems}

A diagonal operator on \lean{lp G 2} or \lean{pi_Lp 2 G} is an operator that, for some fixed sequence of scalars \lstinline{μ : ι → 𝕜}, sends each dependent function \lean{f : Π (i : ι), G i} to the pointwise-rescaled function \lstinline{λ i, μ i • f i}.  The spectral theorem for compact self-adjoint (respectively, normal) operators states that such an operator over \lean{is_R_or_C} (respectively, $\CC$) is equivalent to a diagonal operator on \lean{lp G 2}, for some family of inner product spaces $G$. The finite-dimensional special case, the diagonalization theorem, states that a normal endomorphism of a finite-dimensional inner product space over $\mathbb{C}$ is equivalent to a diagonal operator on some \lean{pi_Lp 2 G}.

The key point of all such theorems, which we defer discussing to  Section~\ref{subsec:exists-eigenvalue},  is a proof that every operator from the stated class has an eigenvalue (unless the operator is the trivial operator on the trivial vector space).  The proof of this important point is what differs from theorem to theorem.  In this subsection we discuss the common part of the proofs of the theorems, namely the reduction to the existence of an eigenvalue.

This part is essentially algebraic and is carried out for a endomorphism of an inner product space $E$ that satisfies the following property, common to those three cases:
\begin{lstlisting}
def inner_product_space.is_normal (T : E →ₗ[𝕜] E) : Prop :=
  ∃ (T' : E →ₗ[𝕜] E), T' * T = T * T' ∧ ∀ x y, ⟪T' x, y⟫ = ⟪x, T y⟫
\end{lstlisting}

We first show that the eigenspaces of such an operator are mutually orthogonal.
\begin{lstlisting}
lemma orthogonal_family_eigenspaces [is_R_or_C 𝕜] [inner_product_space 𝕜 E]
  {T : E →ₗ[𝕜] E} (hT : inner_product_space.is_normal T) :
  orthogonal_family 𝕜 (λ (μ : 𝕜), (eigenspace T μ).subtypeₗᵢ)
\end{lstlisting}  
This puts us in a position to apply the final construction from Section~\ref{subsec:hilbert-sum} to the collection of eigenspaces of $T$.  Specifically, if the completeness property \lstinline{(⨆ (μ : 𝕜), (eigenspace T μ)).topological_closure = ⊤} or its finite-dimensional analogue can be established, then those results establish an isometric isomorphism between $E$ and the Hilbert sum of its own eigenspaces.  It is easy to check that the operator $T$, when transferred by this isometric isomorphism to the Hilbert sum, is diagonal.

A further sequence of lemmas leads to this completeness property, and it is here that the eigenvalue existence result is required.  It is shown that an \lean{inner_product_space.is_normal} operator preserves orthogonal complements of eigenspaces.
\begin{lstlisting}
lemma invariant_orthogonal_eigenspace [is_R_or_C 𝕜] [inner_product_space 𝕜 E]
  {T : E →ₗ[𝕜] E} (hT : inner_product_space.is_normal T) (μ : 𝕜) (v : E)
  (hv : v ∈ (eigenspace T μ)ᗮ) :
  T v ∈ (eigenspace T μ)ᗮ
\end{lstlisting}
Such an operator preserves the mutual orthogonal complement of all its eigenspaces.
\begin{lstlisting}
lemma orthogonal_supr_eigenspaces_invariant [is_R_or_C 𝕜]
  [inner_product_space 𝕜 E] {T : E →ₗ[𝕜] E}
  (hT : inner_product_space.is_normal T) ⦃v : E⦄
  (hv : v ∈ (⨆ (μ : 𝕜), eigenspace T μ)ᗮ) :
  T v ∈ (⨆ (μ : 𝕜), eigenspace T μ)ᗮ
\end{lstlisting}
The restriction of such an operator to this mutual orthogonal complement, which is therefore well-defined, itself has no eigenvalues.
\begin{lstlisting}
lemma orthogonal_supr_eigenspaces [is_R_or_C 𝕜] [inner_product_space 𝕜 E]
  {T : E →ₗ[𝕜] E} (hT : inner_product_space.is_normal T) (μ : 𝕜) :
  eigenspace (T.restrict (orthogonal_supr_eigenspaces_invariant hT)) μ = ⊥
\end{lstlisting}
From here, if the existence of an eigenvalue for all nontrivial operators in the class considered is known, by contraposition the subspace \lean{(⨆ (μ : 𝕜), eigenspace T μ)ᗮ} (being the domain of the operator \lean{T.restrict (orthogonal_supr_eigenspaces_invariant hT)}, which has no eigenvalues) must be trivial.  Standard Hilbert space theory implies that the subspace \lstinline{⨆ (μ : 𝕜), eigenspace T μ} must be dense, the desired completeness result.

\subsection{Existence of an eigenvalue} \label{subsec:exists-eigenvalue}

The first version of the spectral theorem we prove is for normal endomorphisms of a finite-dimensional inner product space over $\CC$.  
\begin{lstlisting}
def diagonalization [inner_product_space ℂ E] [finite_dimensional ℂ E]
  {T : E →ₗ[ℂ] E} (hT : is_star_normal T) :
  E ≃ₗᵢ[ℂ] pi_Lp 2 (λ μ : eigenvalues T, eigenspace T μ)
\end{lstlisting}
    
\begin{lstlisting}
lemma diagonalization_apply_self_apply [inner_product_space ℂ E]
  [finite_dimensional ℂ E] {T : E →ₗ[ℂ] E} (hT : is_star_normal T) (v : E)
  (μ : eigenvalues T) :
  diagonalization hT (T v) μ = (μ : ℂ) • (diagonalization hT) v μ
\end{lstlisting}
We also provide the more classical version of this theorem,
stating that there exists an orthonormal basis of eigenvectors of \lean{T}.

For this class of operators, the proof of the existence of an eigenvalue is  straightforward.  In finite dimension, an endomorphism has a well-defined characteristic polynomial.  Over an algebraically closed field this polynomial must have a root, and this root is an eigenvalue.

The second version of the spectral theorem we prove is for self-adjoint compact operators on a Hilbert space. Here a map between normed spaces is said to be compact, if the image of every bounded subset has compact closure.
\begin{lstlisting}
def compact_map [nondiscrete_normed_field 𝕜] [normed_group E]
  [normed_space 𝕜 E] [normed_group F] (T : E → F) : Prop :=
∀ s : set E, metric.bounded s → is_compact (closure (T '' s))
\end{lstlisting}
A compact linear map is automatically continuous, so it is no loss of generality to take \lean{T} to be of type \lean{E →L[𝕜] E}.
In this setting we state the spectral theorem as follows.
\begin{lstlisting}
def diagonalization' [is_R_or_C 𝕜] [inner_product_space 𝕜 E]
  [complete_space E] {T : E →L[𝕜] E} (hT : T ∈ self_adjoint (E →L[𝕜] E))
  (hT_cpct : compact_map T) :
  E ≃ₗᵢ[𝕜] (lp (λ μ, eigenspace (T : E →ₗ[𝕜] E) μ) 2)
\end{lstlisting}

\begin{lstlisting}
lemma diagonalization_apply_self_apply' [is_R_or_C 𝕜]
  [inner_product_space 𝕜 E] [complete_space E] {T : E →L[𝕜] E}
  (hT : T ∈ self_adjoint (E →L[𝕜] E)) (hT_cpct : compact_map T) (v : E)
  (μ : 𝕜) :
  diagonalization' hT hT_cpct (T v) μ = μ • diagonalization' hT hT_cpct v μ
\end{lstlisting}

For this class of operators, the proof of the existence of an eigenvalue comes from a long and delicate calculation involving the Rayleigh quotient, some 700 lines of code. It is proved that local maxima/minima of the Rayleigh quotient are eigenvectors,
that the operator norm of \lean{T} is the supremum of the absolute value of the Rayleigh quotient, and (using the compactness of \lean{T}) that the Rayleigh quotient of \lean{T} achieves its maximum.

Having established in this project the basic properties of compact operators, the infinite-dimensional theorem of the spectral theorem for compact normal operators is also within reach.  There, the proof of the existence of an eigenvalue comes from an argument about the resolvent, a holomorphic function with values in the Banach space \lstinline{E →ₗ[ℂ] E}.  The current development of complex analysis in \mathlib by Yury Kudryashov~\cite{Kudryashov2022} is sufficiently general for this setting.
However, this would not supersede the spectral theorem we prove for compact self-adjoint operators: the latter works generically over $\mathbb{R}$ and $\mathbb{C}$, which is more elegant than to deduce it in the real setting from the normal-operator version over $\mathbb{C}$ by making an argument about the operator's complexification.

\section{Frobenius-semilinear maps and isocrystals}
\label{sec:isocrystals}

Our formal development of semilinear maps was motivated by 
applications in functional analysis 
to unify statements and proofs over $\RR$ and $\CC$. 
But these maps are interesting and fruitful objects of study in their own right. 
As an example of an interesting result about semilinear maps 
that are \emph{not} linear or conjugate-linear,
we formalize the one-dimensional case of a theorem of Dieudonn\'e and Manin~\cite{Mani63} (see Demazure~\cite[chapter 4]{demazure2006lectures} for a classical exposition and
Lurie~\cite{Lurie2018} for a modern outline without proof),
which classifies the isocrystals over an algebraically closed field of characteristic $p>0$
(Section~\ref{subsec:math-prelim-frobenius}).

We denote the ring of \lean{p}-typical Witt vectors over \lean{k} by \lstinline{𝕎 k} 
and the field of fractions of this ring by \lean{K(p, k)}.
This was defined in \mathlib by Commelin and Lewis~\cite{CL21},
along with the Frobenius endomorphism
\lean{frobenius : 𝕎 k →+* 𝕎 k}.

For the remainder of this section,
we work in a context where 
\lean{p} is a prime natural number
and \lean{k} is an integral domain of characteristic \lean{p}
with a \lean{p}th root function.
\begin{lstlisting}
variables (p : ℕ) [fact p.prime] 
  {k : Type*} [comm_ring k] [is_domain k] [char_p k p] [perfect_ring k p]
\end{lstlisting}
Since the base ring \lean{k} has characteristic \lean{p}, 
\lean{frobenius} satisfies the following property:
\begin{lstlisting}
lemma coeff_frobenius_char_p (x : 𝕎 k) (n : ℕ) :
  (frobenius x).coeff n = (x.coeff n) ^ p
\end{lstlisting}
The additional hypothesis that \lean{k} has a \lean{p}th root function
implies that \lean{frobenius} is in fact an automorphism,
and with \lean{k} an integral domain,
this induces an automorphism on the field of fractions \lean{K(p, k)}.
Locally we let \lstinline{φ(p, k)} denote this map.

We will be interested in maps between \lean{K(p, k)}-vector spaces 
that are semilinear in \lstinline{φ} (``Frobenius-semilinear''). 
To facilitate the use of these maps, 
we add an instance of \lean{ring_hom_inv_pair} (Section~\ref{subsec:composition})
for \lstinline{φ} and its inverse.
We also introduce notation \lean{V →ᶠˡ[p, k] V₂} and \lean{V ≃ᶠˡ[p, k] V₂}
for the types of Frobenius-semilinear maps and equivalences.

An \emph{isocrystal} is a vector space over the field \lean{K(p, k)} 
additionally equipped with a Frobenius-semilinear automorphism. 
\begin{lstlisting}
class isocrystal (V : Type*) [add_comm_group V] extends module K(p, k) V :=
(frob : V ≃ᶠˡ[p, k] V)
\end{lstlisting}
We denote the map \lean{frob} by \lstinline{Φ(p, k)}.
We say two isocrystals over \lean{K(p, k)} are equivalent 
(denoted \lean{V ≃ᶠⁱ[p, k] V₂})
if there is a linear equivalence \lean{f : V ≃ₗ[K(p, k)] V₂} which is ``Frobenius-equivariant,''
that is, for all \lean{x}, \lstinline{Φ(p, k) (f x) = f (Φ(p, k) x)}.

The Dieudonn\'{e}--Manin theorem classifies the isocrystal structures in every finite dimension, up to this notion of equivalence, over an algebraically closed field \lean{k}.  We restrict our attention to the one-dimensional case, where the classification can be stated quite explicitly.  The field \lean{K(p, k)} is naturally a vector space over itself with dimension 1.
There is a standard family of Frobenius-semilinear automorphisms \lean{K(p, k) ≃ᶠˡ[p, k] K(p, k)} indexed by the integers, namely \lean{p^m • φ(p, k)} for each \lean{m : ℤ}, where 
the Frobenius automorphism \lstinline{φ(p, k)} is itself considered as a Frobenius-semilinear automorphism.  
This induces a $\mathbb{Z}$-indexed family of distinct isocrystals 
which we refer to as \lean{standard_one_dim_isocrystal p k m}, and we prove that any one-dimensional isocrystal is 
equivalent to one of the these standard isocrystals.
\begin{lstlisting}
lemma classification [field k] [is_alg_closed k] [char_p k p]
  [add_comm_group V] [isocrystal p k V] (h_dim : finrank K(p, k) V = 1) :
  ∃ (m : ℤ), nonempty (standard_one_dim_isocrystal p k m ≃ᶠⁱ[p, k] V)
\end{lstlisting}

The key to proving this statement 
is finding, for any \lean{a, b :  𝕎 k} with nonzero leading coefficients, 
a vector \lean{x : 𝕎 k} such that \lean{frobenius x * a = x * b}.
We  define such an \lean{x} coefficient by coefficient
by an intricate recursion
that invokes the algebraic closedness of \lean{k} at each step 
to solve a new polynomial equation.
The argument requires us to mediate between different ``levels'' of polynomials---universal multivariate
polynomials over $\ZZ$, and multivariate and univariate polynomials over \lean{k}---which 
proved challenging.
Arithmetic operations on Witt vectors are notoriously complicated,
and the machinery for universal calculations introduced by Commelin and Lewis~\cite{CL21}
does not apply here.
This key lemma takes 550 lines to establish. 

The remainder of the proof of the isocrystal classification theorem was remarkably straightforward. 
We needed to extend \mathlib's existing Witt vector library to show that 
when \lean{k} is an integral domain, \lstinline{𝕎 k} is too.
Modulo this and the key lemma of the previous paragraph,
the proof (including the definitions of Frobenius-semilinear maps and isocrystals)
takes only 100 lines. 

\section{Related work}
\label{sec:conclusion}

Given the fundamental importance of linear algebra,
it is no surprise that theories have been developed in many proof assistants.
To our knowledge, none of these libraries define semilinear maps,
none prove the spectral theorem for compact operators, 
and none prove any of the results we describe 
generically over $\RR$ and $\CC$.

Mahmoud, Aravantinos, and Tahar~\cite{Mahmoud2013}
and Afshar et al.~\cite{Afshar14}
both describe developments in HOL Light of complex vector spaces. 
Both use encodings inherently specific to the complex case; 
they do not generalize the work over the reals by Harrison~\cite{Harrison2013}.

Aransay and Divas\'on~\cite{Aransay15} 
introduce vector spaces over arbitrary fields to Isabelle/HOL, using a careful combination of type classes and Isabelle's \emph{locale} feature. 
A paper by the same authors~\cite{Aransay2017} 
describes an experiment to generalize 
the Isabelle definition of a real inner product space to a larger class of fields, 
 using a type class that seems analogous to our class \lean{is_R_or_C} (Section~\ref{subsec:r-or-c}).
Implementing this idea systematically would probably involve
providing a locale-based generalization of \emph{euclidean-space}
at the beginning of the Isabelle/HOL mathematical analysis library,
and the authors do not take this project on,
despite noting how useful the generalization would be.

An Isabelle Archive of Formal Proofs entry by Caballero and Unruh~\cite{Caballero2021}
duplicates much of the real vector space development
in the complex setting,
in the process introducing conjugate-linear maps and the complex adjoint operator.
Little infrastructure seems to be shared between the real and complex cases.
Their development includes a proof of Fr\'echet--Riesz over $\CC$,
but does not indicate how it might specialize to $\RR$.
Also motivated by applications in quantum computation, 
Bordg et al.~\cite{Bordg2021} define the conjugate-transpose, 
the analogue of the adjoint in the matrix setting,
but again do not generalize to arbitrary fields.

Perhaps related to the more expressive type theory,
Coq developments of linear algebra have taken more advantage of type polymorphism. 
The Mathematical Components library~\cite{Mahb17} features a theory of modules over arbitrary scalar rings,
as does Coquelicot~\cite{Bold15}.
Building on both these libraries, 
MathComp-Analysis~\cite{Affeldt2020} develops structures used in functional analysis,
but does not define anything where specialization to scalars in $\{\RR,\CC\}$ is necessary.

Boldo et al.~\cite{Boldo17} prove 
the real case of Fr\'echet--Riesz using Coquelicot,
on the way to the Lax--Milgram theorem,
but do not address the complex case.
Narita et al.~\cite{Narita2015} do the same in Mizar.
In a branch of the Mathematical Components repository,\footnote{
  \url{https://github.com/math-comp/math-comp/pull/207}
}
Cohen proves the diagonalization theorem for normal matrices.  This is mathematically equivalent to the diagonalization theorem for normal endomorphisms of a finite-dimensional space described at the start of Section~\ref{subsec:exists-eigenvalue}.  Cohen's matrix version could more easily be converted for use in verified numerical analysis, whereas the abstract linear-map version we provide is more convenient in mathematical applications and also admits a more streamlined proof.

\bibliography{mathlib-paper}

\end{document}